# Predicting performance-related properties of refrigerant based on tailored small-molecule functional group contribution

Peilin Cao[1], Ying Geng[1], Nan Feng[1], Xiang Zhang[1], Zhiwen Qi[1], Zhen Song[1,*], Rafiqul Gani[2,3,*]

[1]State Key laboratory of Chemical Engineering, School of Chemical Engineering, East China University of Science and Technology, 130 Meilong Road, Shanghai 200237, China

[2]PSE for SPEED Company, Charlottenlund DK-2920, Denmark

[3]Department of Applied Sustainability, Széchenyi István University, Győr 9026, Hungary

***Corresponding author***: songz@ecust.edu.cn (*Z. S.*), rgani2018@gmail.com (*R. G.*)

**Abstract:** As current group contribution (GC) methods are mostly proposed for a wide size-range of molecules, applying them to property prediction of small refrigerant molecules could lead to unacceptable errors. In this sense, for the design of novel refrigerants and refrigeration systems, tailoring GC-based models specifically fitted to refrigerant molecules is of great interest. In this work, databases of potential refrigerant molecules are first collected, focusing on five key properties related to the operational efficiency of refrigeration systems, namely normal boiling point, critical temperature, critical pressure, enthalpy of vaporization, and acentric factor. Based on tailored small-molecule groups, the GC method is combined with machine learning (ML) to model these performance-related properties. Following the development of GC-ML models, their performance is analyzed to highlight the potential group-to-property contributions. Additionally, the refrigerant property databases are extended internally and externally, based on which examples are presented to highlight the significance of the developed models.

# 1. INTRODUCTION

Refrigerants are extensively utilized in vapor compression refrigeration cycles and are also employed in organic Rankine cycles and solar collectors,[1] which are closely related to the principal global challenges related to energy and environmental issues. However, many refrigerants currently in use have a detrimental impact on the environment including ozone depletion and global warming. Various protection bills, such as the Montreal's Protocol[2,3] and the Kigali Amendment,[4] have been proposed to progressively eliminate refrigerants with adverse environmental impacts. Consequently, identifying environmentally-compatible alternative refrigerants has attracted wide attention of both academia and industry in the refrigeration field.

In the search of alternative refrigerants, subject to the consideration of potential environmental impacts, the primary task is to evaluate the operational efficiency of the corresponding refrigeration systems, which is closely related to several key molecular properties of refrigerants such as critical properties and boiling point.[5] Over the past few decades, researchers have mainly relied on experimental approaches to study the performance-related properties of potential refrigerant molecules and mixtures. However, due to the heavy experimental load and high expenses required, it is impractical to solely apply the experimental approaches for searching promising alternative refrigerant across a large chemical space.[6] In this sense, predictive property models become highly valuable tools to obtain the preliminary estimation of the performance-related properties of refrigerant more efficiently.[7-10]

Several empirical correlations relying primarily on easily-measurable physical properties such as molecular weight and density have been proposed for simple and fast calculations of target refrigerant properties,[11] but they are criticized for the lack of theoretical basis and limited universality.[12] Another widely used method is equation of state (EoS),[13,14] which has a favorable thermodynamic interpretation. However, the parameters of EoSs are usually regressed from experimental data, leading to the lack of universal applicability on unstudied systems.[5] In the context of computer aided molecular design (CAMD),[15–18] a more pragmatic method is the quantitative structure-property relationship (QSPR) models. There are diverse structural representation approaches in QSPR modelling such as groups, descriptors or fingerprints.[19,20] Besides, learned representations from molecular sequence and graph have also been popularly used very recently in some models like convolutional neural networks[21] and graph neural networks.[22] Compared with the other representation approaches that are not easy to support model interpretation, dividing molecules into their constituent groups is the simplest and most straightforward method and the resultant group contribution (GC) models have easy-to-interpret meaning.[23-25] Notably, GC models uniquely benefit both the forward development of property modelling and the reverse mathematical synthesis of novel molecules, leveraging well-established rules between molecular structure feasibility/complexity and groups.[26]

When evaluating current GC models and functional group decomposition approaches, one could find that most refrigerant molecules are either poorly tackled or it is not possible to tackle them. The underlying reason is that current GC models are mostly developed for a wide size-range of molecules by defining relatively large functional groups, whereas refrigerant molecules are mainly small molecules with different functional characteristics.[27,28] In fact, many refrigerant molecules cannot be represented by applying any of the currently available functional group decomposition schemes based on large molecules. Moreover, it should be pointed out that the prediction error of refrigerant property, such as the normal boiling point, should preferably be less than 1%.[29] Otherwise, there would be unacceptable discrepancy between the predicted and the actual temperature of a designed refrigeration system.[30] Hence, GC-based models specifically tailored to small refrigerant molecules with acceptable prediction accuracy are highly desirable. To this end, tailoring suitable small-molecule functional group decomposition scheme is one prerequisite and pairing a powerful modelling algorithm is another. In recent years, machine learning (ML) has gained prominence for accurately estimating molecular properties.[31-34] Though the ML-based models are usually "black-box" and not easy to integrate with molecular design algorithms,[33] the synergy between ML and GC presents a promising pathway to greatly improve both the accuracy and interpretability for molecular property prediction.[35–38]

Taking all the aforementioned aspects into account, this work first collects the most comprehensive refrigerant database covering five key performance-related properties of refrigerant namely normal boiling point ($T_b$), critical pressure ($P_c$), critical temperature ($T_c$), enthalpy of vaporization at 298 K (Hv[298K]), and acentric factor ($\omega$). A tailor-made GC method is proposed to specifically represent small refrigerant molecules for model development. Four different ML algorithms namely random forest (RF), extreme gradient boosting (XGB), support vector regression (SVR), and gaussian process regression (GPR) are tested for the purpose of developing a high-accuracy model.

The remainder of the paper is organized as follows. Following the introduction, theoretical backgrounds on GC method, ML algorithms for model development, and SHAP (Shapely additive explanation) analysis for model interpretation are briefly described in Section 2. In Section 3, the analysis of the collected data for the five performance-related refrigerant properties and the tailored small-molecule GC method are given together with model development details. Section 4 presents the results of GC-based modelling and the interpretation analysis of the final models. In Section 5, the final models are employed for internal and external extension of the property database, and subsequently case studies based on the extended databases are presented. Conclusions are summarized in Section 6.

## 2. THEORETICAL BACKGROUND

### 2.1 GC method

In classical GC-simple modelling, the molecular structure of a component is first decomposed into building blocks, known as functional groups (to be called groups). The essence of GC methods lies in the assumption that the contribution of a specific group to a certain molecular property is the same across all molecules containing it. Then, based on the additive principle, the property of a molecule is determined by the summation of the contributions of its constituent groups. The GC method applied in this work is similar to the Marrero-Gani (MG) method,[39] which can be expressed as:

$$f(x) = \sum_i N_i C_i + \sum_j M_j D_j + \sum_k O_k E_k \tag{1}$$

where $f(x)$ is a function of the property $x$ to be modelled; $C_i$ represents the contribution of the atoms of type-$i$ that occurs $N_i$ times in a molecule; $D_j$ denotes the contribution of the bonds of type-$j$ that occurs $M_j$ times in a molecule; $E_k$ is the contribution of the first-, second- and third-order groups of type-$k$ that occurs $O_k$ times in a molecule.

### 2.2 ML algorithms

The ML algorithms are used to regress the model coefficients, that is, the descriptor (groups, atoms, bonds) contributions for each property. Note that these contributions can be one value or a complex nonlinear mapping. Four algorithms are described briefly that are tested to develop the GC-ML models.

**2.2.1 RF algorithm**

Random Forest (RF)[40] is a highly effective ensemble learning algorithm that uses decision tree (DT)[41] as its base learner. As a nonparametric ML algorithm, DT tackles prediction tasks by deriving decision rules from model training and presenting the rules in a treelike structure. By aggregating the predictions of multiple randomized DTs, the stability and accuracy of the model is improved by RF based on the bagging idea (a contraction of bootstrap-aggregating). Assume the mean-squared generalization error of DT as:

$$E_{x,y} = (y - h(x))^2 \quad (2)$$

where $E$ means the mean-squared error; $y$ is the value of label; $h(x)$ is the predicted value of a single regression tree. RF is a combination of several randomized DTs and aggregates their predictions by averaging, giving the overall mean-squared generalization error as:

$$E_{forest} = \lim_{n=\infty} E_{X,Y}\left(Y - \frac{h(X,\theta_n)}{n}\right)^2 = E_{X,Y}(Y - E_\theta h(X,\theta))^2 \quad (3)$$

where $\theta$ denotes the randomly split vector to grow the trees, aiming at reducing the average error of prediction to increase the performance of RF. The prediction of RF can be expressed as:

$$y_{pre} = \frac{1}{N}\sum_1^N h_i(x) \quad (4)$$

where $h_i(x)$ denotes the prediction of the $i$th tree. In contrast to the DT approach that evaluates all features to determine the best pruning, RF operates by randomly choosing a subset of features and finding the best trees for them to avoid overfitting and improve model performance.

**2.2.2 XGB algorithm**

XGB, as proposed by Chen and Guestrin[42] based on the gradient boosting decision tree (GBDT) framework,[43] is a widely recognized and well-established scalable ML algorithm. It employs an iterative ensemble approach where each tree builds upon the previous one. In XGB modelling, the initial tree is trained to predict the target, while the subsequent tree is trained to predict the residuals between the predictions of the first tree and the real value. Each new tree aims to address the remaining residuals from the previous trees, enhancing the overall prediction accuracy. XGB iteratively repeat the process to achieve the objective of boosting, as presented by:

$$Loss = \sum_{i=1}^{n} l(\hat{y}_i, y_i) + \sum_{k}^{N} \Omega(f_k) \tag{5}$$

where $l$ represents a differentiable convex loss function, $\Omega$ denotes the regularization term, and $f_k$ refers to the $k$th tree. The prediction of XGB can be expressed as:

$$y_{pre} = \sum_{1}^{K} f_k(x) \tag{6}$$

Typically, gradient descent is guided by the first-order derivative of the loss function. XGB enhances this approach by calculating the second-order derivative of the loss function, which accounts for the curvature of the gradient. Compared to GBDT, this refinement leads to a faster training and improved prediction accuracy.

**2.2.3 SVR algorithm**

SVR is a widely utilized nonlinear regression algorithm known for its stability and robustness. Unlike linear algorithms, SVR employs kernel methods to transform input data into higher-dimensional spaces, replacing the standard dot product with kernel estimation to handle complex relationships.[44] For a given set of input vectors $x$ and the corresponding label vectors $y$, the developed linear algorithm in SVR can be expressed

as follows:

$$y_{pre} = w^T x + b = b + \sum_{i=1}^{n} \alpha_i x^T x_i \qquad (7)$$

where $\alpha_i$ denotes the coefficient vector, $w$ is the weight of features, $b$ is the bias of the function. The prediction function of nonlinear regression can be expressed by replacing the dot product with a kernel estimation as follows:

$$k(x, x_i) = \varphi(x) \cdot \varphi(x_i) \qquad (8)$$

$$y_{pre} = b + \sum_{i=1}^{n} \alpha_i k(x, x_i) \qquad (9)$$

where $k$ refers to the kernel function. The newly obtained function is linear for $\alpha$ and $\varphi(x)$ but nonlinear for $x$. SVR transforms the input data into a higher-dimensional space through the function $\varphi(x)$, allowing the application of a new linear model in this space. Therefore, SVR highlights the critical importance of selecting the appropriate kernel function, as it directly affects the performance. Options for kernel function include RBF kernel, polynomial kernel, sigmoid kernel, ANOVA kernel, among others.

**2.2.4 GPR algorithm**

GPR[45] is a Bayesian regression method that excels in function assumption and uncertainty estimation. The GPR method models a probability distribution over the values of a function evaluated at a finite set of input vectors, denoted as $n_1, \ldots, n_M$, which assumes that the joint distribution $p(f(n_1), \ldots, f(n_M))$ follows a Gaussian distribution. The core objective of GPR model is to predict the property $y$ of molecules using a target function, which can be expressed as:

$$y_{pre} = K_{1 \times M} W_{M \times 1} \qquad (10)$$

where $K$ means the matrix composed of the values calculated by the kernel function, and $W$ is the coefficients learned from the training data. Meanwhile, GPR model can indicate the prediction uncertainty by the covariance calculated by:

$$Cov_{1\times 1} = K_{1\times 1} - K_{1\times M}U_{M\times M}K_{M\times 1} \quad (11)$$

where $U$ means the noise covariance representing the uncertainty of the training data. The performance of the GPR model exclusively relies on the kernel function, which controls the shape and characteristics of the fitted function. To effectively capture complex patterns and trends from the data, GPR can flexibly combine multiple kernel functions. For instance, GPR can utilize a combination of linear, polynomial, and RBF kernels to calculate $K$, thereby promoting its predictive ability.

**2.3 SHAP interpretation**

Model interpretation is crucial for investigating how ML models generate specific predictions, which is key to developing accurate and dependable predictive models. In this work, SHAP[46] analysis is employed to interpret the outputs of ML models, which is rooted in a cooperative game theory. It constructs an additive explanatory model to analyze how predictions change when each feature is excluded, treating the features as contributors and aggregating these effects to compute SHAP values. The SHAP value is calculated according to:

$$\hat{\Phi}_j = \frac{1}{K}\sum_{K=1}^{K}\left(\hat{g}(x_{+j}^m) - \hat{g}(x_{-j}^m)\right) \quad (12)$$

where $\hat{\Phi}_j$ is the SHAP value for feature $j$, $\hat{g}(x_{+j}^m)$ is the prediction for $x$. The model calculates SHAP values for every predicted sample, which measures the contribution of each feature to the prediction for one specific sample. In this context, the SHAP values reflect both the significance of each feature and its positive or negative influence on predictions, thereby showing the interpretation of complex models.

## 3. GC-MODEL DEVELOPMENT

### 3.1. Database collection and analysis

For modelling refrigerant properties, a database of small molecules that could be potentially used as refrigerants is first collected from ASHRAE[47] and NIST.[48] A total of 1,766 chemicals are classified into different types, including hydrocarbons, chlorinated compounds, fluorinated compounds, and others. The number of compounds and the corresponding lowest and highest molecular weights for each type are listed in Table S1 (supporting material document). Considering the five performance-related properties, measured data of molecules having 1 - 7 carbon atoms in the database are utilized for modelling. As listed in Table 1, there are 1016, 321, 336, 158 and 227 data points for normal boiling point ($T_b$), critical temperature ($T_c$), critical pressure ($P_c$), heat of vaporization at 298K (Hv[298K]), and acentric factor ($\omega$), respectively. Figure 1 shows the data distribution of the five property datasets, where 90% of the data points are distributed in the range of 260.5 – 437.3 K for $T_b$, 376.5 – 630.8 K for $T_c$, 25.9 – 55.5 bar for $P_c$, 23.7 – 47.6 kJ/mol for Hv[298K], and 0.153 - 0.375 for $\omega$ (as indicated by the dashed lines).

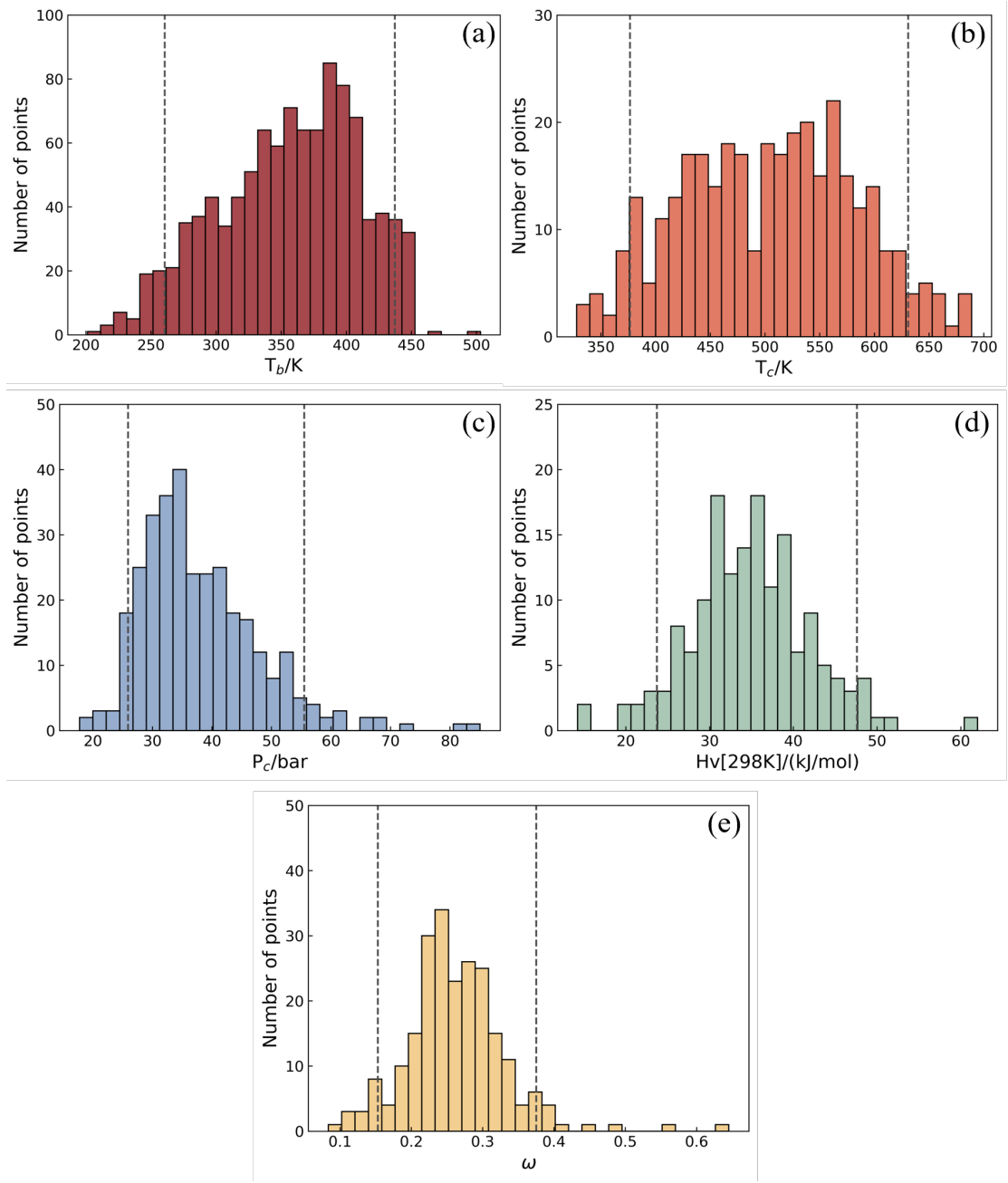

**Figure 1.** Data distribution of each property dataset: (a) $T_b$; (b) $T_c$; (c) $P_c$; (d) Hv[298K]; (e) $\omega$.

**Table 1.** Information of the five property datasets.

|  | $T_b$/K | $T_c$/K | $P_c$/bar | Hv[298K]/ (kJ/mol) | $\omega$ |
|---|---|---|---|---|---|
| Data points | 1016 | 336 | 321 | 158 | 227 |
| Groups | 231 | 154 | 151 | 78 | 123 |
| Data range | 201.0-503.0 | 327.8-689.0 | 17.8-85.0 | 14.2-62.0 | 0.08-0.64 |

### 3.2. Small-molecule groups

In this work, a new set of 231 small-molecule descriptors consisting of first-order groups, special groups, atoms and bonds are used to represent the refrigerant molecules in the database. The full list of descriptors is given as Table S2 (supporting material document) and three examples of representation of molecules with descriptors are shown in Figure 2. Figure 2a shows the descriptors representing a single C-atom molecule ($CH_2F_2$) where atoms (in this case are also the first-order groups), bonds and special groups are used. Figures 2b and 2c show the descriptors for two and three C-atoms molecules ($CH_2Cl_2F_2$ and $C_3H_2F_4$) where first-order and special (second-order) groups are used. The representations for all the molecules in the database for $T_b$, $T_c$, $P_c$, $Hv[298K]$, and $\omega$ are given in Supplementary Material (in GitHub). As the numbers of compounds differ for each property, the numbers of descriptors used for each property also differ (231, 154, 151, 78, and 123, respectively for $T_b$, $T_c$, $P_c$, $Hv[298K]$, and $\omega$). Note that, the rules for representation of molecules by first- and higher-order groups reported by Marrero and Gani[39] and Hukkeriker et al.[49] are employed here for molecules with two or more C-atoms. For single C-atom molecules, the first-order groups (that is, atoms) satisfy the valency rules, while the bonds and special groups defined for small molecules are used as additional descriptors. These descriptors cover all compounds with one C-atom connected with H, Cl, F, Br atoms. Also, for two and higher C-atoms molecules, special second-order groups have been defined.

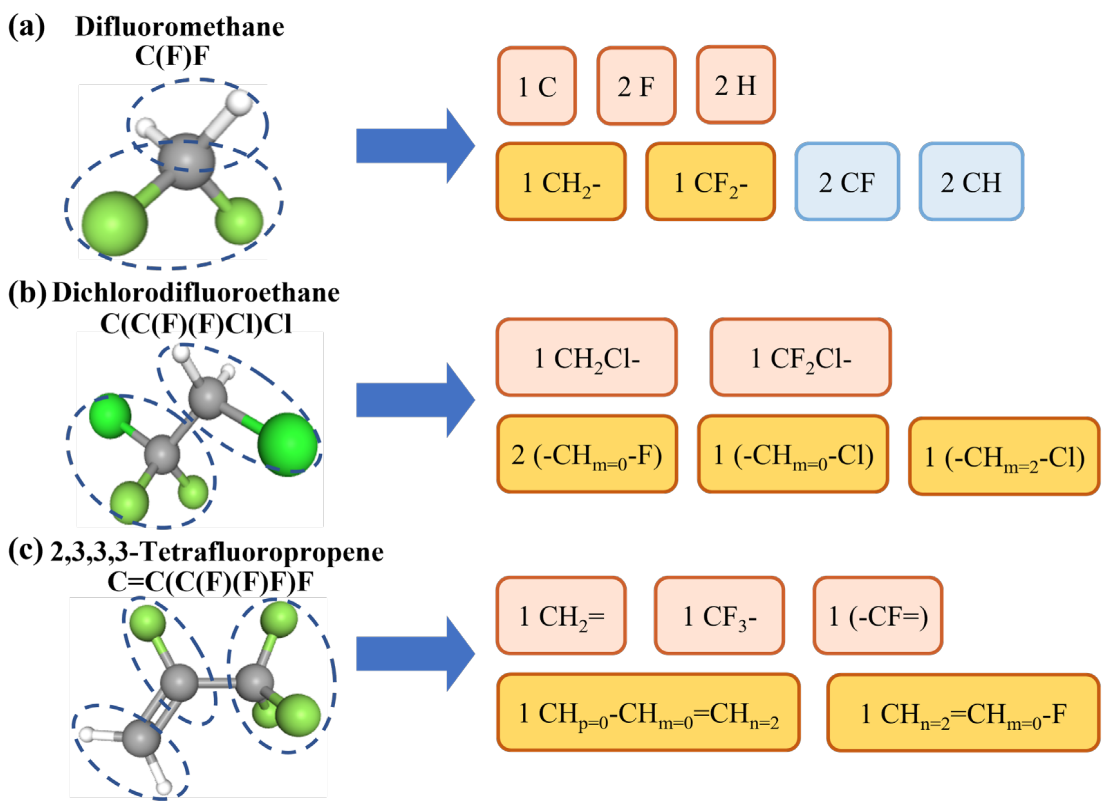

**Figure 2.** Examples of decomposing molecules to the tailored small-molecule groups (light red indicates atoms or first-order groups; blue indicates bonds for single C-atom molecules; orange indicates special or second-order groups).

### 3.3. Model development

**3.3.1 GC-ML model**

At the start of developing the GC-ML model, ten disjoint subsets are obtained by randomly dividing the initial dataset. Following that, each subset is sequentially used as a test set, while the remaining nine subsets serve as the training set. To optimize the hyperparameters of the GC-ML models, a five-fold cross-validation procedure is implemented with a grid search approach in model training. The detailed space of hyperparameters for the optimization of each model is summarized in Table S3 (supporting material document). Once the optimal hyperparameters are determined, the entire training dataset is utilized to build a model, which is then evaluated on the corresponding test subset. Finally, the predictions of all ten test subsets are combined to provide a comprehensive assessment of the overall performance of the ML algorithm. To evaluate the performance of GC-ML models, two statistic metrics namely the correlation coefficient ($R^2$) and the mean absolute error (MAE) are used. Figure 3 shows the framework for the development of GC-ML model. The algorithm for the final model of each property is selected through the comprehensive comparison on the four ML algorithms RF, XGB, SVR, and GPR. These algorithms and the corresponding optimization process are implemented with the Scikit-learn package.[50]

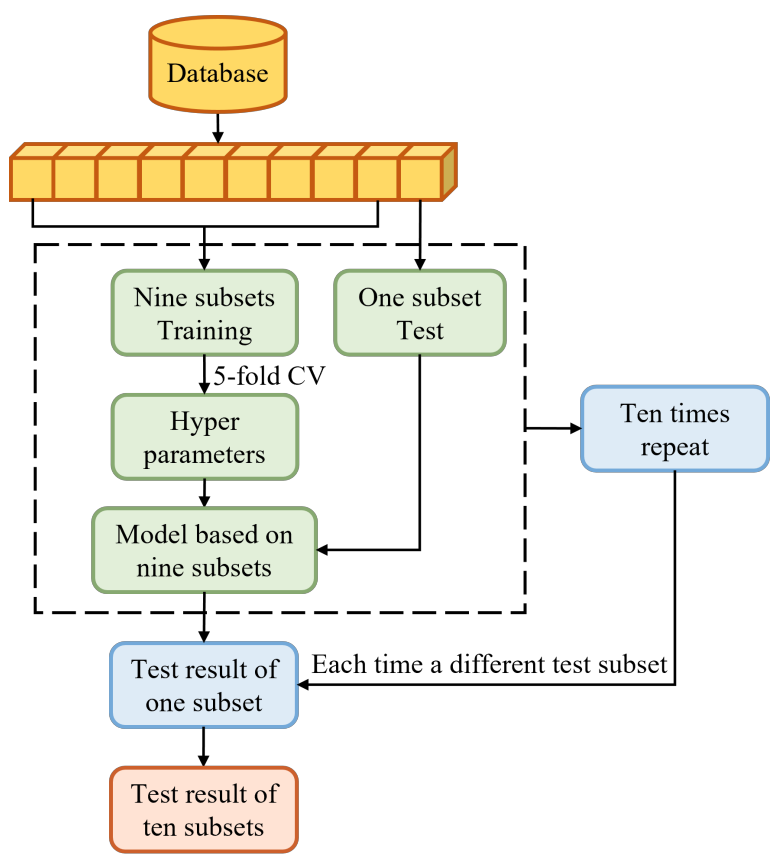

**Figure 3.** Framework for the development of GC-ML model.

### 3.3.2 GC-simple model and empirical correlation model

For comparison of different models for molecular property modelling, $T_b$, $T_c$, and $P_c$ are exemplified to investigate the potential of GC-simple model (for $T_b$ and $P_c$) and empirical correlation model (for $T_c$). In this work, the parameters of GC-simple model are regressed utilizing the stepwise regression method previously employed by Hukkerikar et al.[49,51] The optimization process focuses on minimizing the squared error function, which quantifies the error between the measured and estimated property values, as expressed by:

$$S(N, M, O) = \min w_k \sum_{j=1}^{N}(y_j - y_{pre,j})^2 \qquad (13)$$

where $w_k$ is a parameter that can be adjusted during both sequential and simultaneous regression of the GC-simple model. A maximum likelihood-based regression method is employed to minimize the squared error. For $T_c$ modelling, an empirical correlation model is developed to estimate $T_c$ from measured $T_b$ values, as presented:

$$T_c = P_1 + (P_2 MW) + (P_3 T_b) + (P_4 T_b^{P_5}) \qquad (14)$$

where $P_1$ - $P_5$ are parameters used in the empirical correlation and MW is the molecular weight. Subsequently, the results of these models can be evaluated and compared with those of the GC-ML models to illustrate the performance of different models on modelling the refrigerant properties.

## 4. RESULTS AND DISCUSSION

### 4.1. GC-ML modelling results

Following the modelling procedure introduced in Figure 3, Table 2 records the results of two statistic metrics ($R^2$, MAE) in training and testing the GC-ML models. Comparing both the training and test results, it can be found that GPR outperforms the other three algorithms on modelling all the five properties. To be more specific, the test $R^2$ of SVR and GPR are higher than those of RF and XGB, suggesting that the algorithms based on kernel function are better than those based on trees for the GC-ML modelling of the five datasets. The advantages of GPR primarily lie in the flexible choice of kernel functions and the strong capability for uncertainty estimation, resulting in better performance than that of the other three algorithms in the GC-ML modelling. The parity plot between experimental and the GC-GPR model tested values for the five properties is presented in Figure S1 (supporting material document), showing the reasonable test performance of the model. Consequently, GPR is selected for the development of the final model for further applications on refrigerant screening and design.

**Table 2**. $R^2$ and MAE of training and test of GC-ML modeling.

| | $T_b$/K | | | | $T_c$/K | | | | $P_c$/bar | | | | Hv[298K]/(kJ/mol) | | | | $\omega$% | | | |
|---|---|---|---|---|---|---|---|---|---|---|---|---|---|---|---|---|---|---|---|---|
| | Training | | Test | | Training | | Test | | Training | | Test | | Training | | Test | | Training | | Test | |
| | $R^2$ | MAE | $R^2$ | MAE | $R^2$ | MAE | $R^2$ | MAE | $R^2$ | MAE | $R^2$ | MAE | $R^2$ | MAE | $R^2$ | MAE | $R^2$ | MAE | $R^2$ | MAE |
| RF | 0.956 | 8.132 | 0.684 | 21.938 | 0.950 | 13.097 | 0.683 | 33.470 | 0.946 | 1.426 | 0.641 | 3.653 | 0.919 | 1.437 | 0.489 | 3.687 | 0.897 | 0.057 | 0.314 | 0.147 |
| XGB | 0.983 | 4.518 | 0.761 | 17.941 | 0.932 | 15.146 | 0.711 | 31.641 | 0.982 | 0.880 | 0.643 | 3.605 | 0.977 | 0.631 | 0.515 | 3.641 | 0.881 | 0.059 | 0.245 | 0.154 |
| SVR | 0.996 | 0.708 | 0.857 | 11.606 | 0.980 | 3.649 | 0.786 | 23.193 | 0.972 | 0.467 | 0.740 | 2.687 | 0.905 | 0.856 | 0.616 | 2.846 | 0.897 | 0.052 | 0.454 | 0.136 |
| **GPR** | **0.999** | **0.284** | **0.882** | **9.671** | **0.999** | **0.534** | **0.788** | **22.012** | **0.999** | **0.048** | **0.749** | **2.699** | **0.999** | **0.033** | **0.621** | **2.877** | **0.886** | **0.058** | **0.468** | **0.135** |

## 4.2 Performance of the final model

### 4.2.1 Performance of GC-GPR model

Employing GPR as the algorithm for the final model, the GC-GPR model is trained on the whole dataset to evaluate the performance. Figure 4 shows the ARD as a function of compound number (ordered in terms of increasing ARD), wherein the ARD of 992, 327, 290, 152 and 199 compounds are less than 1% for $T_b$, $T_c$, $P_c$, $H_v$[298K] and $\omega$, respectively. As observed, nearly all compounds have less than 5% error in the modelling of the five performance-related properties, indicating that the GC-ML models are powerful alternatives to the GC-simple model and correlation model from quantitative modeling point of view due to an extended model complexity. To more vividly show the performance of the final model, the parity plot between the experimental and estimated values is presented in Figure 5. As shown with the diagonal as a reference, the centralized distribution of data points visibly illustrates the encouraging performance of the final models for all the five properties. In this context, the final models could be considered reliable for the applications on refrigerant screening and design.

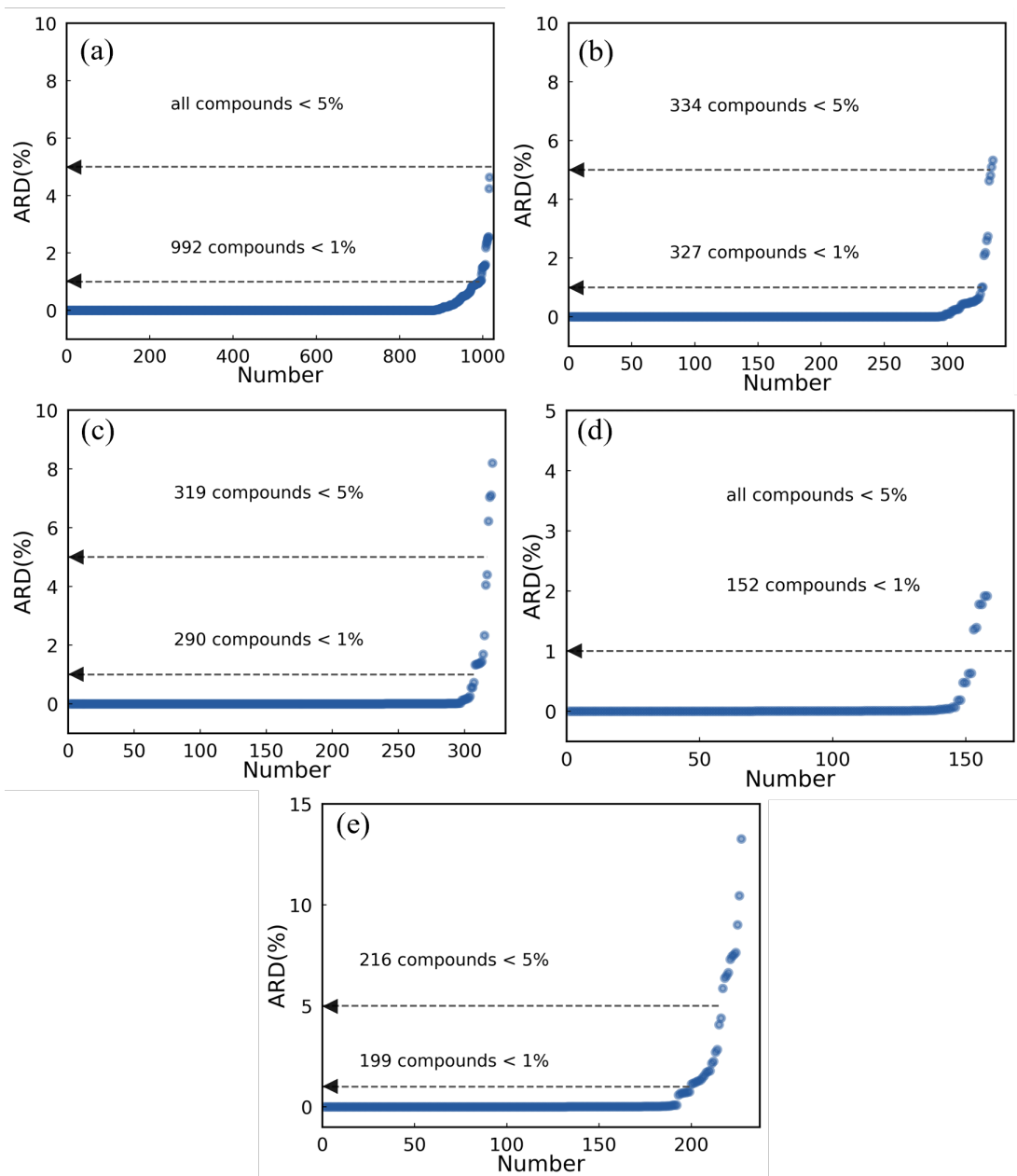

**Figure 4.** ARD as a function of compound number: (a) $T_b$; (b) $T_c$; (c) $P_c$; (d) $H_v[298K]$; (e) $\omega$.

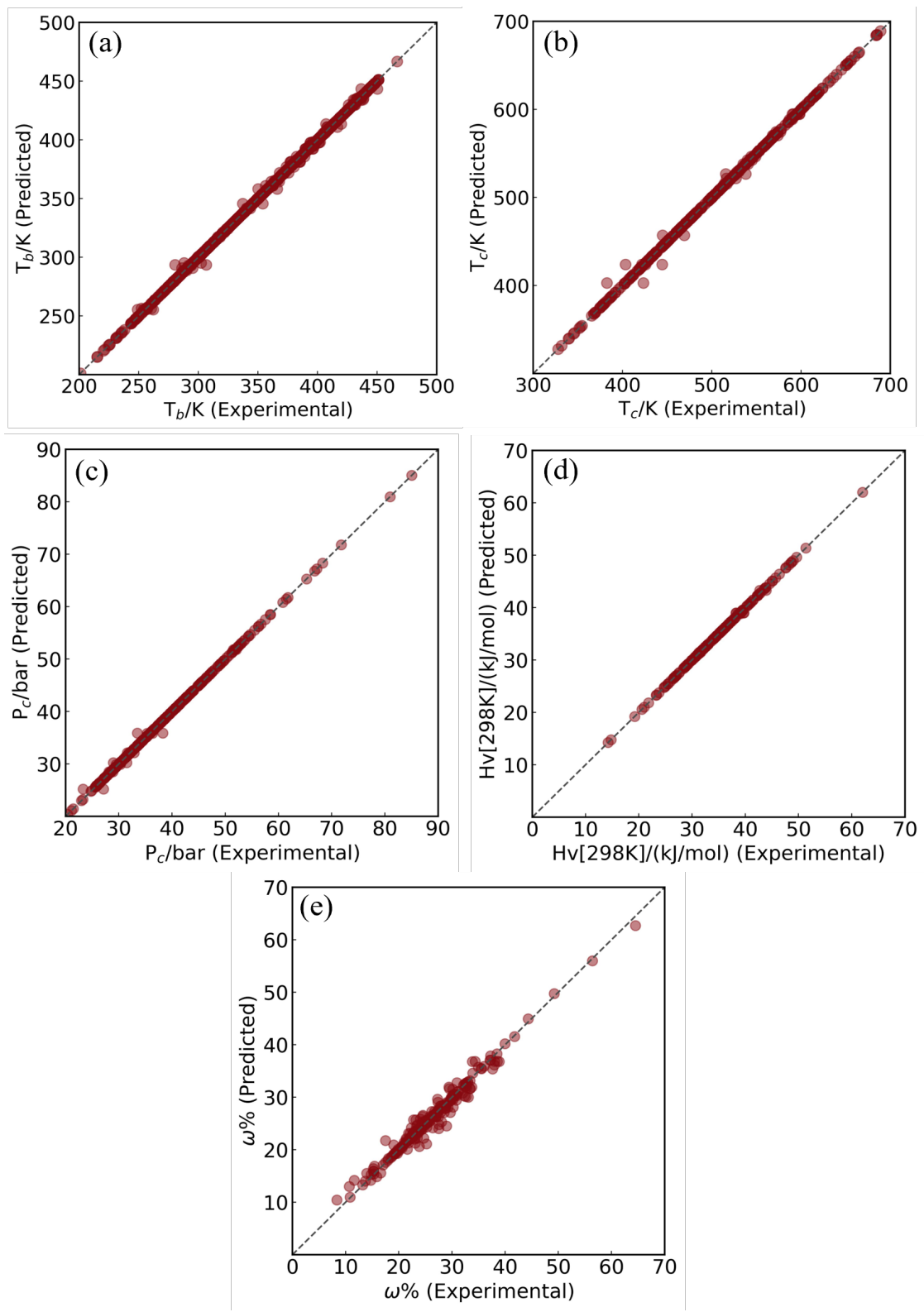

**Figure 5.** Plots of experimental versus GC-GPR model estimated values: (a) $T_b$; (b) $T_c$; (c) $P_c$; (d) Hv[298K]; (e) ω.

**4.2.2 Comparison on different models**

In this section, the performance of the GC-simple model and the empirical correlation model are presented. According to the results shown in Table 3, the ARD of most of compounds are less than 10%, showing a reasonable estimation of the GC-simple model and the correlation model. However, only about 90% compounds have less than 5% error and a small proportion of compounds has less than 1% error in modelling the three properties, which is yet not satisfactory for refrigeration applications. To be more specific, the ARD of 40.4%, 21.1%, and 34.0% of compounds are less than 1% for modelling $P_c$, $T_b$, and $T_c$, respectively. In comparison, over 90% of compounds have less than 1% ARD in modelling the three properties of GC-GPR model, obviously showing that the performance of the GC-based model can be further improved by employing ML algorithms. Since achieving high quantitative prediction accuracy is critical for effective refrigerant screening and design, GC-ML models could offer great advantages in the modelling of performance-related refrigerant properties.

**Table 3.** The distribution of ARD for estimated properties compared with measured data.

|  | GC simple model | | $T_b$-$T_c$ relation | GC-GPR model | | |
| --- | --- | --- | --- | --- | --- | --- |
|  | $P_c$/bar | $T_b$/K | $T_c$/K | $P_c$/bar | $T_b$/K | $T_c$/K |
| <1% | 34.0 | 40.4 | 21.1 | 90.3 | 97.6 | 97.3 |
| <5% | 83.2 | 91.8 | 89.0 | 99.4 | 100.0 | 99.4 |
| <10% | 97.2 | 99.3 | 98.5 | 100.0 | 100.0 | 100.0 |

**4.3 SHAP analysis**

To interpret the outputs of the final models, SHAP analysis is performed to investigate the influence of features on the properties. The global feature importance map of SHAP analysis is presented in Figure 6, where features with larger mean absolute SHAP value have greater influence on model. It can be observed that (-CH2-) has the greatest impact on modelling $T_b$, $T_c$, $H_v[298K]$ and $\omega$, whereas CH3- has the greatest impact on modelling $P_c$. The mean absolute SHAP values of all features for each property are listed in Supplementary Material (in GitHub). To further present the positive and negative effects of the features, the bee colony map of SHAP analysis is shown in Figure S2 (supporting material document). Using the (-CH2-) feature in Figure S2a as an example, the bee colony map illustrates that as the value of the (-CH2-) feature increases, the color of the points shifts from blue to red and their SHAP values change from negative to positive. The analysis indicates that a higher number of (-CH2-) groups in a refrigerant is associated with a higher predicted $T_b$ by the model. In a similar manner, the SHAP values of most features to $P_c$ are negative and those to $T_b$, $T_c$, $H_v[298K]$ and $\omega$ are positive, indicating that a higher molecular weight of a refrigerant is associated with a lower $P_c$ and higher $T_b$, $T_c$, $H_v[298K]$ and $\omega$. The concrete SHAP analysis on model interpretation reveal the impact of group features on the estimation of the five properties, thereby indicating the effective insights of the model on the small-molecule groups.

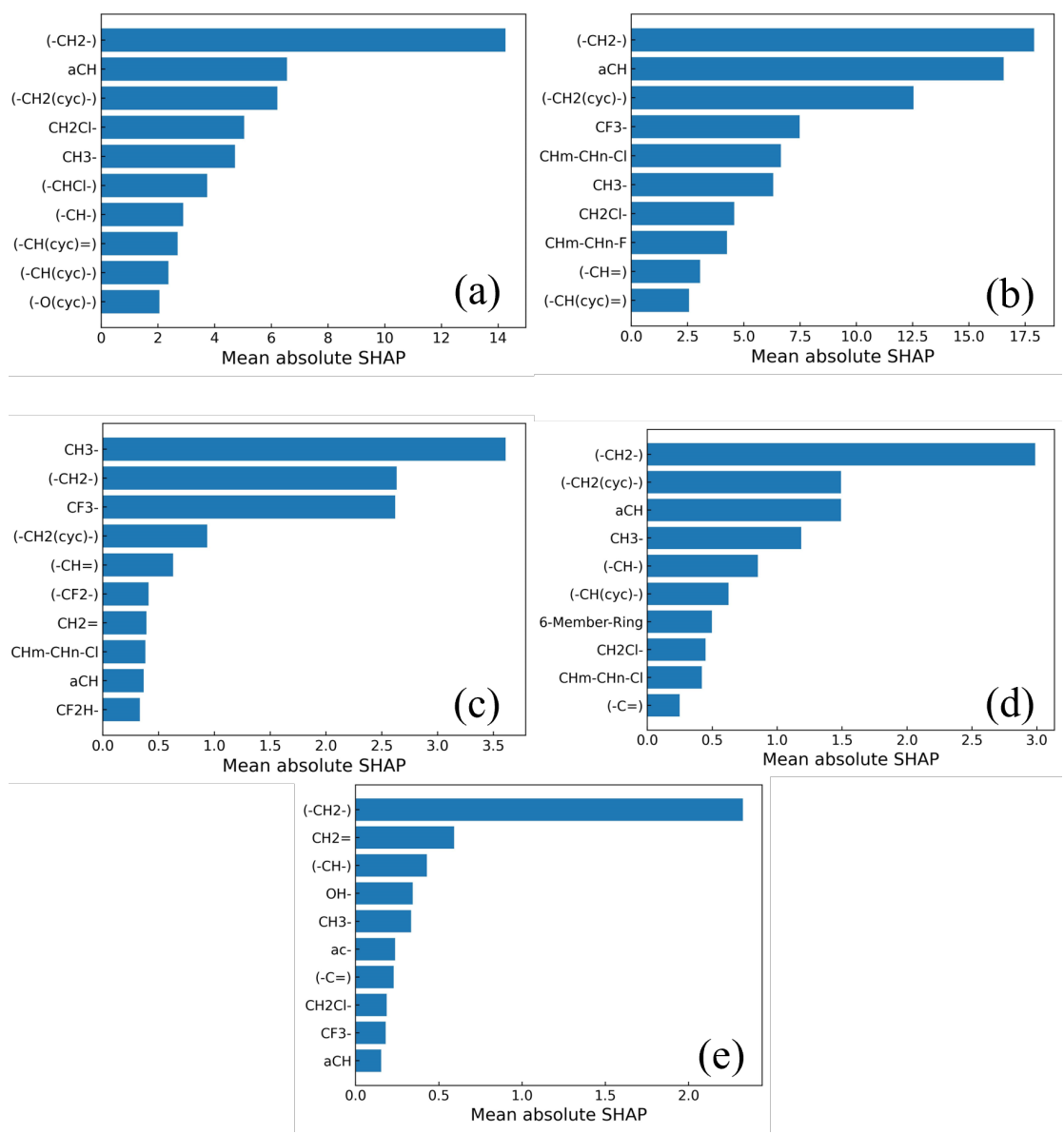

**Figure 6.** Global feature importance map from SHAP analysis on the final models: (a) $T_b$; (b) $T_c$; (c) $P_c$; (d) $H_v[298K]$; (e) ω.

## 5. MODEL APPLICATION

### 5.1. Internal database extension

Given that the number of molecules in $T_b$ is the largest among those in the five datasets, the final models are first applied for estimating the blank measured values of $T_c$, $P_c$, $H_v[298K]$ and ω on the 1016 molecules (referred to as the "internal database extension"). According to the flowsheet shown in Figure 7, among the molecules that can be represented by the current group vectors of the corresponding property models, 534, 530, 346 and 545 molecules can be estimated in the internal database extension for $T_c$, $P_c$, Hv[298K] and $\omega$, respectively. Figure S3 (supporting material document) shows the data distributions of internal extension, where the ranges of the estimated value are 375.8 - 648.2 K, 17.8 - 60.9 bar, 23.0 - 46.6 kJ/mol, and 0.20 - 0.66 for $T_c$, $P_c$, Hv[298K] and $\omega$, respectively. The data distributions of the internal extension are close to those of the original property datasets (listed in Table 1), indicating the reliability of the model prediction and the potential of the internal extended database for refrigerant screening.

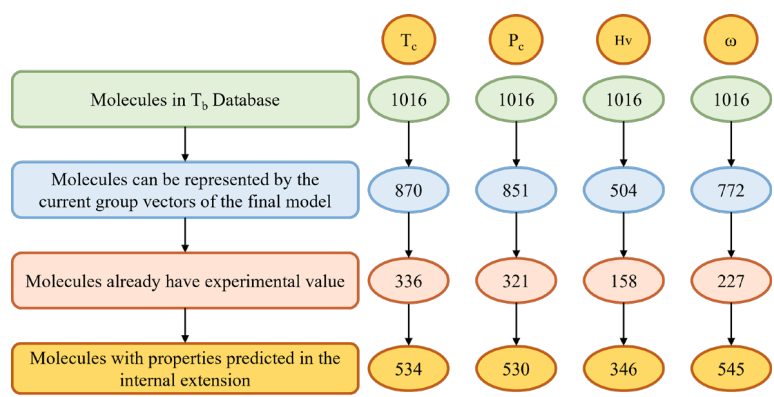

**Figure 7.** Flowsheet for searching molecules without measured values of $T_c$, $P_c$, Hv[298K], and $\omega$ in the internal database extension.

**5.2. External database extension**

With the application of the final models, the properties of unreported potential refrigerants can also be estimated (referred to as the "external database extension"). By substituting one of the small-molecule groups in the one-carbon and two-carbon refrigerant molecules from the original datasets, 306, 265, 241, 48, and 170 molecules are obtained for the external database extension of $T_b$, $T_c$, $P_c$, Hv[298K] and $\omega$, respectively (as listed in Table 4). These molecules as well as the corresponding constituent groups are recorded in the Supplementary Material (in GitHub), which are consequently used as the input of the final model to estimate the properties. Figure S4 (supporting material document) shows the data distribution for the estimated properties of these molecules, which are in the range of 220.4 - 506.1 K, 348.9 - 646.8 K, 31.6 - 70.9 bar, 26.0 - 56.4 kJ/mol, and 0.20 - 0.43 for $T_b$, $T_c$, $P_c$, $H_v[298K]$ and $\omega$, respectively. Notably, provided that a novel molecule can be decomposed by the defined small-molecule groups, the relevant performance-related property can be predicted by the final model. Combing results of the internal and external extension, the extended property database can be a valuable library for screening refrigerants to help improve the operational efficiency of the refrigeration systems.

Table 4. Information of the external database extension.

|  | $T_b$/K | $T_c$/K | $P_c$/bar | $H_v[298K]$/(kJ/mol) | $\omega$ |
| --- | --- | --- | --- | --- | --- |
| Data points | 306 | 265 | 241 | 48 | 170 |
| Range | 220.4-506.1 | 348.9-646.8 | 31.6-70.9 | 26.0-56.4 | 0.20-0.43 |

**5.3. Case study on molecules with similar structure**

Prior to utilizing the extended database to screen refrigerants, a case study on molecules with similar structure (that is, ClCH2F, Cl(CH2)2F, Cl(CH2)3F, Cl(CH2)4F, and Cl(CH2)5F) is presented in Figure 8. As observed, the incorporation of the (-CH2-) group to molecules gives rise to an increase in $T_b$ and $T_c$ (represented by the red and orange points, respectively) and a decrease in $P_c$ (represented by the blue points). For instance, the $T_c$ increases form 471.15 K to 566.13 K with the gradual addition of the (-CH2-) group to the compound, which is consistent with the SHAP analysis and again demonstrates its impact on the three properties. The consistency between the analysis on similar molecular structures and on model interpretation indicates both the effective understanding of the GC-ML model on the performance-related properties and the reliability of the extended database for refrigerant screening. It should be noted that though the chemical formulas of some molecules are similar, the structure of them as well as the corresponding performance-related properties are distinguished. For instance, the triangle points (representing $C_4H_8F_2$) are at a distance on the y-axis from the circle points (representing $C_4H_8FCl$) of the same color on the dashed line shown in Figure 8. The discrepancy in properties between $C_4H_8F_2$ and $C_4H_8FCl$ demonstrates that a comparable chemical formula does not necessarily imply similar molecular properties. Thus, the identification of alternative molecules for refrigerants should focus on the molecular properties rather than relying on simple similarity of chemical formula.

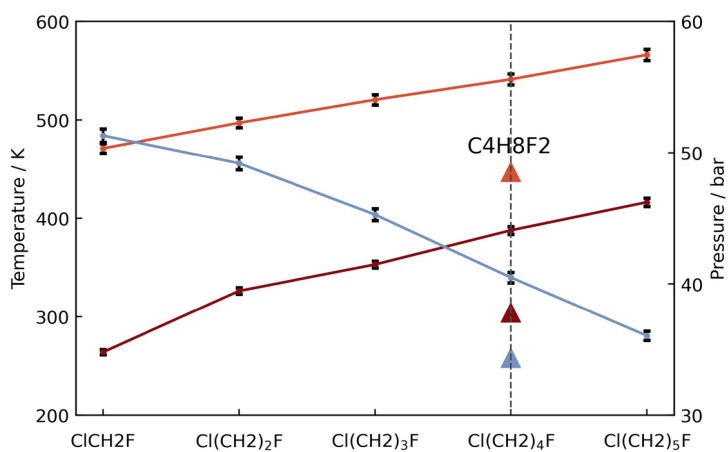

**Figure 8.** Case study on molecules with similar structure: red points for $T_b$; orange points for $T_c$; blue points for $P_c$; the triangles for properties of $C_4H_8F_2$.

**5.4. Case study on screening alternatives of existing refrigerants**

By searching the extended database, the molecules with available $T_b$, $T_c$, and $P_c$ (either measured or estimated) are integrated for screening potential alternative molecules for reference refrigerant (R-134a in this case). Figure 9 vividly shows the distribution of molecules with the three properties in the extended database, where the red star denotes R-134a, the circle represents the data points from the internal extension, and the triangle represents the data points from the external extension. Table 5 lists molecules with the three properties simultaneously within ±5% deviation from those of the reference refrigerant. For example, the $T_b$, $T_c$, and $P_c$ of 1,1-Difluoropropene are 244.12 K, 382.00 K and 40.64 bar, which are in the range of 234.27 - 258.93 K, 355.59 - 393.02 K, and 38.61 - 42.67 bar, respectively (±5% deviation from the properties of R-134a). It should be noted that unreported potential molecules such as 2,2,2-Trifluoroethylamine could also serve as alternatives to existing refrigerants. Hence, in the event that a refrigeration system seeks alternative refrigerants or has a requirement for refrigerants with properties in specified ranges, the extended property database in this work can be a valuable resource for identifying suitable candidates to help improving the operational efficiency.

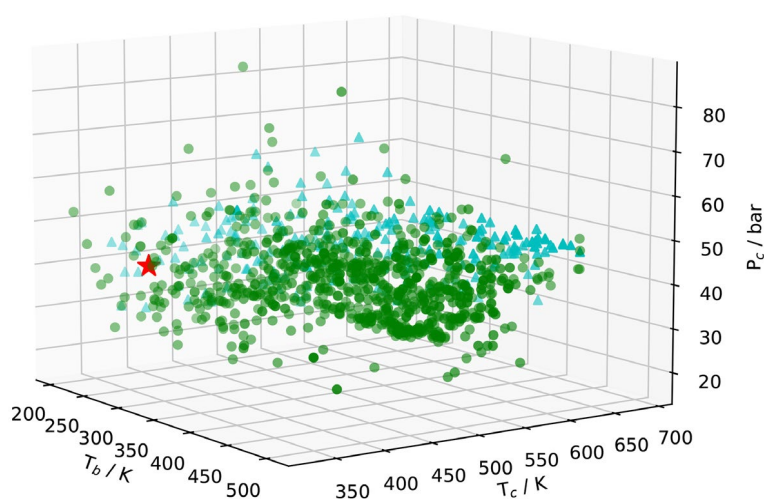

**Figure 9.** Data distribution of compounds with the three properties ($T_b$, $T_c$, and $P_c$) in the extended database.

Table 5. Example of alternatives of R-134a refrigerant.

| Component | $P_c$/bar | $T_b$/K | $T_c$/K |
|---|---|---|---|
| R-134a (Reference) | 40.64* | 246.60* | 374.30* |
| 1,1-Difluoropropene | 42.22* | 244.12* | 382.00* |
| Dichlorodifluoromethane | 41.25* | 243.35* | 384.95* |
| Chlorotrifluoroethylene | 40.53* | 245.35* | 379.15* |
| Tetrafluoroethylene | 39.37 | 242.76 | 365.83 |
| 2,2,2-Trifluoroethylamine | 42.06 | 252.17 | 381.89 |

*denotes the measured property values of the molecule in the original database.

## 6. CONCLUSION

In this work, the GC-ML based refrigerant property modelling is presented by tailoring small-molecule group decomposition, which covers five refrigeration performance-related properties (normal boiling point, critical pressure and temperature, enthalpy of vaporization at 298 K, and acentric factor). Compared to the GC-simple model and empirical correlation model, the GC-ML models can produce superior performance. Particularly, GPR performs better than the other three ML algorithms, with 97.6%, 97.3%, 90.3%, 96.2%, and 87.7% of compounds having less than 1% ARD in modelling $T_b$, $T_c$, $P_c$, $H_v[298K]$ and $\omega$, respectively. Furthermore, SHAP analysis reveals the impact of the group features on each final model, showing that (-CH2-) has the greatest impact on modelling $T_b$, $T_c$, $H_v[298K]$ and $\omega$, whereas CH3- has the greatest impact on modelling $P_c$. The insights from the SHAP analysis indicates that a higher molecular weight of a refrigerant is associated with a lower $P_c$ and higher $T_b$, $T_c$, $H_v[298K]$ and $\omega$. Following that, based on the database extended by the application of the final models, the case study on molecules with similar structure shows consistency with the SHAP analysis. An example of screening potential alternative refrigerants to R134a is presented by searching the extended database, where 1,1-Difluoropropene, Dichlorodifluoromethane, Chlorotrifluoroethylene, Tetrafluoroethylene, and 2,2,2-Trifluoroethylamine are identified. Notably, beyond these five performance-related properties, the tailored small-molecule GC method could be further extended for the modelling of environmental, health and safety properties of refrigerants. The resultant models in this manner could be integrated as valuable tools for the computer-aided screening or design of novel refrigerants as well as refrigeration cycle simulation toward the improvement of refrigeration systems.

## ACKNOWLEDGEMENTS

This research is supported by the National Natural Science Foundation of China (NSFC) under the grants of 22208098 and 22278134. Z. S also acknowledges the support by the Fundamental Research Funds for the Central Universities under the grant of JKA01231663.

## DATA AVAILABILITY AND REPRODUCIBILITY STATEMENT

The data and final models that support the findings of this study are openly available (as Supplementary Materials) in GitHub at https://github.com/Ecust-PeilinCao/Refrigerant-property-modelling.git. For Figures 1, 8, 9, S3, and S4, the original and extended databases with the group composition of molecules are available in the Supplementary Material (in the given GitHub repository). The numerical data of SHAP values for Figures 6 and S2 are tabulated in the Supplementary Material (in the given GitHub repository). For Figures 4, 5, and S1, the predicted property values of all molecules in this work are available in the Supplementary Material (in the given GitHub repository). Error bars (where shown) in Figure 8 show the ±1% deviation from the measured refrigerant properties.

## SUPPORTING MATERIAL DOCUMENT

Additional supporting information can be found online in a separate supporting material document (classification of the collected small molecules in Table S1, new small-molecule groups of this work in Table S2, search space of hyperparameters for each GC-ML model in Table S3, plots of experimental versus tested values by GC-GPR model in Figure S1, bee colony map from SHAP analysis on the final models in Figure S2, data distribution for the internal and external database extension in Figures S3 and S4, contents of data and models tabulated in the Supplementary Material) that can be downloaded from the publisher.

## REFERENCES

1. Mondejar ME, Haglind F. The potential of halogenated olefins as working fluids for organic Rankine cycle technology. *J Mol Liq*. 2020;310:112971.

2. Montreal Protocol. *Article 2F: Hydrochlorofluorocarbons.* UN Environmental Program, Ozone Secretariat.

3. Montreal Protocol. *Article 2A: CFCs*. UN Environmental Program, Ozone Secretariat.

4. Montreal Protocol. *The Kigali Amendment (2016): The amendment to the Montreal Protocol agreed by the Twenty-Eighth Meeting of the Parties (Kigali, 10−15 October 2016).* UN Environmental Program, Ozone Secretariat.

5. Alkhatib III, Albà CG, Darwish AS, Llovell F, Vega LF. Searching for sustainable refrigerants by bridging molecular modeling with machine learning. *Ind Eng Chem Res*. 2022;61(21):7414-7429.

6. Su W, Zhao L, Deng S. Group contribution methods in thermodynamic cycles: Physical properties estimation of pure working fluids. *Renew Sust Energ Rev*. 2017;79:984-1001.

7. Ferro VR, Ruiz E, Tobajas M, Palomar JF. Integration of COSMO‐based methodologies into commercial process simulators: Separation and purification of reuterin. *AIChE J*. 2012;58(11):3404-3415.

8. Yu G, Wei Z, Chen K, Guo R, Lei Z. Predictive molecular thermodynamic models for ionic liquids. *AIChE J*. 2022;68(4):e17575.

9. Shi, H., Zhou, T. Computational design of heterogeneous catalysts and gas separation materials for advanced chemical processing. *Front Chem Sci Eng*. 2021;15(1):49-59.

10. Yin H, Xu M, Luo Z, et al. Machine learning for membrane design and discovery. *Green Energy Environ*. 2024;9(1):54-70.

11. Vejahati F, Nikoo MB, Mokhatab S, Towler BF. Simple correlation estimates critical properties of alkanes. *Pet Sci Technol*. 2007;25(9):1115-1123.

12. Que Y, Ren S, Hu Z, Ren J. Machine learning prediction of critical temperature of organic refrigerants by molecular topology. *Processes*. 2022;10(3):577.

13. Gao N, Chen G, Tang L. A corresponding state equation for the prediction of isobaric heat capacity of liquid HFC and HFO refrigerants. *Fluid Phase Equilib*. 2018;456:1-6.

14. Zang L, Zhu Q, Yun Z. Critical properties prediction based on a quartic equation of state. *Can J Chem Eng.* 2010;88(6):1003-1009.

15. Alshehri AS, Gani R, You F. Deep learning and knowledge-based methods for computer-aided molecular design—toward a unified approach: State-of-the-art and future directions. *Comput Chem Eng*. 2020;141:107005.

16. Gani R, Nielsen B, Fredenslund A. A group contribution approach to computer-aided molecular design. *AIChE J*. 1991;37(9):1318-1332.

17. Harper PM, Gani R. A multi-step and multi-level approach for computer aided molecular design. *Comput Chem Eng*. 2000;24(2-7):677-683.

18. Decardi-Nelson B, Alshehri AS, You F. Generative artificial intelligence in chemical engineering spans multiple scales. *Front Chem Eng*. 2024;6:1458156.

19. Song Z, Chen J, Cheng J, Chen G, Qi Z. Computer-aided molecular design of ionic liquids as advanced process media: A review from fundamentals to applications. *Chem Rev*. 2024;124(2):248-317.

20. Zhang L, Mao H, Liu L, Du J, Gani R. A machine learning based computer-aided molecular design/screening methodology for fragrance molecules. *Comput Chem Eng*. 2018;115:295-308.

21. Chen G, Song Z, Qi Z, Sundmacher K. Generalizing property prediction of ionic liquids from limited labeled data: a one-stop framework empowered by transfer learning. *Digital Discov.* 2023;2(3):591-601.

22. Qiu Y, Chen J, Xie K, Gu R, Qi Z, Song Z. Graph transformer based transfer learning for aqueous pK prediction of organic small molecules. *Chem Eng Sci*. 2024;300:120559.

23. Conte E, Martinho A, Matos HA, Gani R. Combined group-contribution and atom connectivity index-based methods for estimation of surface tension and viscosity. *Ind Eng Chem Res*. 2008;47(20):7940-7954.

24. Rueben L, Schilling J, Rehner P, et al. Predicting the relative static permittivity: A group contribution method based on perturbation theory. *J Chem Eng Data.*

2024;69(2):414-426.

25. Chen Y, Ma S, Lei Y, et al. Ionic liquid binary mixtures: Machine learning‐assisted modeling, solvent tailoring, process design, and optimization. *AIChE J.* 2024:e18392.

26. Cao P, Chen J, Chen G, Qi Z, Song Z. A critical methodological revisit on group-contribution based property prediction of ionic liquids with machine learning. *Chem Eng Sci.* 2024;298:120395.

27. Alshehri AS, Tula AK, You F, Gani R. Next generation pure component property estimation models: With and without machine learning techniques. *AIChE J.* 2022;68(6).

28. Cao X, Gong M, Tula A, Chen X, Gani R, Venkatasubramanian V. An improved machine learning model for pure component property estimation. *Engineering.* 2024;39:61-73.

29. Kuprasertwong N, Padungwatanaroj O, Robin A, et al. Computer-aided refrigerant design: new developments. In: Computer aided chemical engineering. Vol 50. *Elsevier*; 2021:19-24.

30. Duvedi AP, Achenie LEK. Designing environmentally safe refrigerants using mathematical programming. *Chem Eng Sci.* 1996;51(15):3727-3739.

31. Li K, Zhu Y, Shi S, et al. Machine learning models coupled with ionic fragment σ-profiles to predict ammonia solubility in ionic liquids. *Green Chem Eng.* 2024:S2666952824000608.

32. Deng Y, Eden M, Cremaschi S. A Gaussian process embedded feature selection method based on automatic relevance determination. *Comput Chem Eng.* 2024;191:108852.

33. Zhang K, Mann V, Venkatasubramanian V. G‐MATT: Single‐step retrosynthesis

prediction using molecular grammar tree transformer. *AIChE J*. 2024;70(1):e18244.

34. Chen Z, Chen J, Qiu Y, et al. Prediction of electrical conductivity of ionic liquids: from COSMO-RS derived QSPR evaluation to boosting machine learning. *ACS Sustainable Chem Eng*. 2024;12(17):6648-6658.

35. Gani R. Group contribution-based property estimation methods: advances and perspectives. *Curr Opin Chem Eng*. 2019;23:184-196.

36. Mondejar ME, Cignitti S, Abildskov J, Woodley JM, Haglind F. Prediction of properties of new halogenated olefins using two group contribution approaches. *Fluid Phase Equilib*. 2017;433:79-96.

37. Devotta S, Chelani A. Unified artificial neural network-group contribution method for predictions of normal boiling point and critical temperature of refrigerants and related compounds. *Int J Refrigeration*. 2022;140:112-124.

38. Liu B, Karimi Nouroddin M. Application of artificial intelligent approach to predict the normal boiling point of refrigerants. *Int J Refrigeration*. 2023;2023:1-9.

39. Marrero J, Gani R. Group-contribution based estimation of pure component properties. *Fluid Phase Equilib*. 2001;183-184:183-208.

40. Breiman, L. Random forests. *Mach Learn.* 2001, 45, 5-32.

41. Quinlan JR. Learning decision tree classifiers. *ACM Comput Surv*. 1996;28(1):71-72.

42. Cao DS, Xu QS, Liang YZ, Chen X, Li HD. Automatic feature subset selection for decision tree-based ensemble methods in the prediction of bioactivity. *Chemometr Intell Lab Syst*. 2010;103(2):129-136.

43. Chen T, Guestrin C. XGBoost: A scalable tree boosting system. In: Proceedings of the 22nd ACM SIGKDD International Conference on knowledge discovery and data

mining. *ACM*; 2016:785-794.

44. Zhou ZH. Machine learning. *Springer Nature*; 2021.

45. Quiñonero-Candela J, Rasmussen CE. A unifying view of sparse approximate Gaussian process regression. *J Mach Learn Res.* 2005; 6(Dec):1939-1959.

46. Lundberg SM, Lee SI. A unified approach to interpreting model predictions. *Adv Neural Inf Process Syst*. 2017;3:30-39.

47. ASHRAE Refrigerant Designations. ANSI/ASHRAE 34-2013, Designation and Safety Classification of Refrigerants, 46-51.

48. The National Institute of Standards and Technology (NIST), 2018, NIST Standard Reference Database Number 69, https://webbook.nist.gov/chemistry/.

49. Hukkerikar AS, Sarup B, Ten Kate A, Abildskov J, Sin G, Gani R. Group-contribution+ (GC+) based estimation of properties of pure components: Improved property estimation and uncertainty analysis. *Fluid Phase Equilib.* 2012;321:25-43.

50. Pedregosa F, Varoquaux G, Gramfort A. Scikit-learn: machine learning in python. *J Mach Learn Res*. 2011;12:2825-2830.

51. Hukkerikar AS, Kalakul S, Sarup B, Young DM, Sin G, Gani R. Estimation of environment-related properties of chemicals for design of sustainable processes: development of group-contribution+ (GC+) property models and uncertainty analysis. *J Chem Inf Model*. 2012;52(11):2823-2839.